# Range dependent Hamiltonian Algorithm for numerical QUBO formulation

Hyunju Lee[a], Kyungtaek Jun[a,*]

[a] Research Center, Innovative Quantum Computed Tomography, Seoul, Republic of Korea

[*] ktfriends@gmail.com

## Abstract

With the advent and development of quantum computers, various quantum algorithms that can solve linear equations and eigenvalues faster than classical computers have been developed. The Harrow-Hassidim-Lloyd algorithm is an algorithm that can solve linear equations in a gate model quantum computer. Still, it is constrained by the use of quantum RAM and the size limit of the matrix according to the total number of qubits in the quantum computer. Recently, Jun and Lee developed a QUBO model for solving linear systems and eigenvalue problems in the quantum computer. However, even though their model uses 2048 qubits, the number of qubits for variables that can be used for the problem is only 64. To solve this problem, we introduce an algorithm that can be used by dividing the size of the entire domain according to the number of qubits. We also form a QUBO model related to each subregion.

## Introduction

Quantum computing has completely changed the computing paradigm.[1] Quantum computing is based on a computation that harnesses various quantum states, such as superposition, entanglement, and interference. A quantum computer had the potential to simulate and calculate things a classical computer could not achieve.[2] Currently, quantum computing is affecting most business and research fields since its various optimistic applications. It is also known for some problems that quantum computing is much faster than classical computing. The most representative examples are the Shor's algorithm[3] as the factorization algorithm and the Grover's algorithm[4] as a quantum search algorithm. The Shor's algorithm can provide exponential speed up and the Grover's algorithm provides a quadratic speed up over their classical counterparts. As another example, Lloyd presented a quantum algorithm to predict the features of solution of linear equations.[5]

Linear equations play a crucial role in many fields of science and engineering. However, the data constituting the linear equations has gradually increased, and then the need for a method that can solve large-scale linear equations more quickly is emerging. As an alternative to this problem, several attempts have been made to solve linear equations in quantum computers. In 2009, Harrow, Hassidim, and Lloyd proposed a monumental quantum algorithm for solving linear systems on gate model quantum computers.[6] Since then, various methods for solving linear equations have been developed, but there is only a solution for a specific type of matrix. Recently, Jun and Lee proposed the numerical QUBO formulation methods that can solve linear systems and eigenvalue problems with $n$ by $n$ general matrix in the quantum annealer.[7] This method has an advantage of speedup because parallel computing is possible in the QUBO modeling process. Unfortunately, due to the limitation of the number of the number of qubits in the current quantum computer, it is difficult to directly apply to the problem of a matrix of too large size.

In this paper, in order to solve this problem, we present a method that can solve the problem of a matrix with a large size than the existing ones for the currently available number of qubits. It is a QUBO formulation by dividing a given domain range into subranges. Using this method, it is possible to solve linear equations and eigenvalue problems for matrices with a larger size with a given available qubits. In this paper, we show the test results for a matrix of size 16 by 16 and 32 by 32 in D-Wave 2000Q.

## Method

### Background

Quadratic unconstrained binary optimization (QUBO) is a combinatorial optimization problem with the wide range as finance and economics[8], for many problems from computer science, and embedding into QUBO has been formulated.[9] To solve a problem using the quantum annealer, we must express it as Ising function defined on logical variables or an equivalent QUBO. Then we can embed the logical problem in the physical architecture of the quantum annealer by mapping the logical variables and qubits. Since quantum annealing algorithms are based on quantum effects to local minima of a cost function by the effect of tunneling, we can find the global minimum of a cost function.[10] In this problem, a cost function $f$ is defined on an $n$-dimensional binary vector space $\mathbb{B}^n$ onto $\mathbb{R}$.

$$f(\vec{x}) = \vec{x}^T Q \vec{x} \qquad \text{Equation 1}$$

where $Q$ is an upper diagonal matrix, $\vec{x} = (x_1, \cdots, x_N)^T$, and $x_i \in \{0,1\}$. The problem is to find $x^*$, which minimizes the cost function $f$. Then finding $x^*$ is equivalent to finding the ground state of the corresponding QUBO formulation,

$$f(\vec{x}) = \sum_{i=1}^{N} Q_{i,i} x_i + \sum_{i<j}^{N} Q_{i,j} x_i x_j. \qquad \text{Equation 2}$$

In $Q$, the diagonal terms $Q_{i,i}$ means the linear terms and the off-diagonal terms $Q_{i,j}$ means the quadratic terms.

For given $A \in \mathbb{R}^{n \times n}$, $\vec{x} \in \mathbb{R}^n$, and $\vec{b} \in \mathbb{R}^n$, the linear least squares problem is to find the $\vec{x}$ that minimizes $\| A\vec{x} - \vec{b} \|$. To solve this problem let us begin by writing out $\| A\vec{x} - \vec{b} \|$:

$$A\vec{x} - \vec{b} = \begin{pmatrix} a_{1,1} & a_{1,2} & \cdots & a_{1,n} \\ a_{2,1} & a_{2,2} & \cdots & a_{2,n} \\ \vdots & \vdots & \ddots & \vdots \\ a_{n,1} & a_{n,2} & \cdots & a_{n,n} \end{pmatrix} \begin{pmatrix} x_1 \\ x_2 \\ \vdots \\ x_n \end{pmatrix} - \begin{pmatrix} b_1 \\ b_2 \\ \vdots \\ b_n \end{pmatrix} \qquad \text{Equation 3}$$

Taking the 2-norm square of the resultant vector of Eq. 3, we get the following:

$$\| A\vec{x} - \vec{b} \|_2^2 = \vec{x}^T A^T A \vec{x} - 2\vec{b}^T A \vec{x} + \vec{b}^T \vec{b} \qquad \text{Equation 4}$$

$$= \sum_{k=1}^{n} \left( \sum_{i=1}^{n} (a_{k,i} x_i)^2 + 2 \sum_{i<j} a_{k,i} a_{k,j} x_i x_j - 2 b_k \sum_{i=1}^{n} a_{k,i} x_i + b_k^2 \right) \qquad \text{Equation 5}$$

Now, as from the study of O'Malley and Vesselinov[11], each $x_i$ could be expressed by the combination of qubits $q_{i,l} \in \{0,1\}$ follows:

$$x_i \approx \sum_{l=-m}^{m} 2^l q_{i,l} \qquad \text{Equation 6}$$

where the integer and fractional parts of $x_i$ are positive $l$ and negative $l$, respectively.

Substituting Eq. 6 into Eq. 5, using $q^2 = q$, each term in Eq. 5 is represented as a combination of linear terms and quadratic terms as follows:

$$\sum_{k=1}^{n} \sum_{i=1}^{n} (a_{k,i} x_i)^2 \approx \sum_{k=1}^{n} \sum_{i=1}^{n} \sum_{l=-m}^{m} a_{k,i}^2 2^{2l} q_{i,l}^+ + \sum_{k=1}^{n} \sum_{i=1}^{n} \sum_{l_1 < l_2} a_{k,i}^2 2^{l_1+l_2+1} q_{i,l_1}^+ q_{i,l_2}^+ \qquad \text{Equation 7}$$

$$\sum_{k=1}^{n}\sum_{i<j} 2\,a_{k,i}a_{k,j}x_i x_j \approx \sum_{k=1}^{n}\sum_{i<j}\sum_{l_1=-m}^{m}\sum_{l_2=-m}^{m} 2^{l_1+l_2+1}\,a_{k,i}a_{k,j}q_{i,l_1}^{+}q_{j,l_2}^{+} \qquad \text{Equation 8}$$

$$\sum_{k=1}^{n}\sum_{i=1}^{n}(-2 a_{k,i}b_k x_i) \approx \sum_{k=1}^{n}\sum_{i=1}^{n}\sum_{l=-m}^{m} 2^{l+1}\,a_{k,i}b_k(-q_{i,l}^{+}) \qquad \text{Equation 9}$$

If the QUBO formulation created in the above process is calculated by quantum annealer, we will obtain the minimum value of the cost function and find $x^*$.

## QUBO modeling for Subranges

Let's look at the effect of adequately dividing the domain range into subranges. First, each $x_i$ can be expressed in the same way as in Eq. 6 as follows:

$$x_i \approx \sum_{l=0}^{m} 2^l\, q_{i,l} \qquad \text{Equation 10}$$

However, this equation, unlike Equation 6, is that $x_i$ can only represent positive integers. So, we will add a translation coefficient for each subrange to represent the positive real number. Now, we can represent subrange dependent $x_i'$ as below

$$x_i' \approx \sum_{l=0}^{m} 2^l\, q_{i,l} + T_{i,s} \qquad \text{Equation 11}$$

where $T_{i,s} = c2^{m+1}$ is translation number for indicating subrange. If the total range of $x_i$ is $(-s2^{m+1}, s2^{m+1}-1)$, the translation coefficient number $c$ is $-s \leq c \leq s-1$.

Let's look at an example to explain it more easily. We consider a linear system for a 2 by 2 matrix to describe the subrange. For each $x_i$, we assume that the total range is from -64 to 63 and let us limit the number of qubits to 4 (See in Fig. 1). In here, since $x_i' = \sum_{l=0}^{3} 2^l\, q_l$ can represent any integer from 0 to 15, the interval can be divided by a multiple of 16.

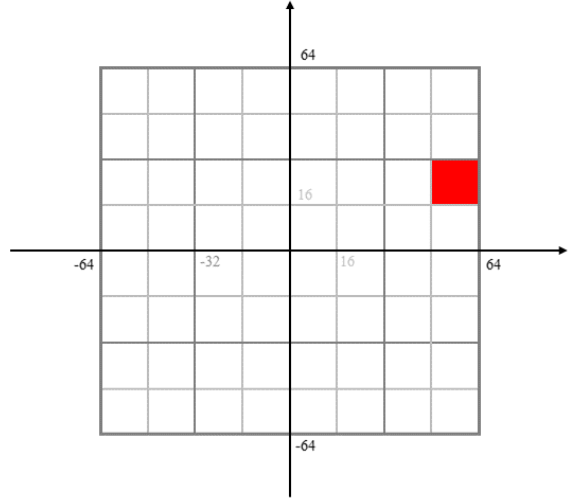

Figure 1. Figure shows how to divide the range from -64 to 63 into subranges when each $x_i$ uses 4 qubits for a 2*2 matrix. The red box means that the one subregion in which the sums of qubits can change in $x_i$ when $m = 3$.

For calculating a QUBO formula, let $\vec{c} = \vec{b} - A\vec{T}$ where a vector $\vec{T} = (T_{1,s}, T_{2,s}, \cdots T_{n,s})^T$. When each $x'_i$ of a vector $\vec{x'}$ is represented by $x'_i \approx \sum_{l=0}^{m} 2^l q_{i,l} + T_{i,s}$, the QUBO formulation can be calculated as follows:

$$\left\|A\vec{x'} - \vec{b}\right\|_2^2 = \left\|A\vec{x} + A\vec{T} - \vec{b}\right\|_2^2 = \left\|A\vec{x} - (\vec{b} - A\vec{T})\right\|_2^2 \qquad \text{Equation 12}$$

$$= \|A\vec{x} - \vec{c}\|_2^2 \qquad \text{Equation 13}$$

$$= \vec{x}^T A^T A \vec{x} - 2\vec{c}^T A\vec{x} + \vec{c}^T \vec{c} \qquad \text{Equation 14}$$

Thus, a minimum value is obtained by QUBO formulation for each subarea, and one subregion containing the solution can have a minimum value of $-\vec{c}^T\vec{c}$.

**Python codes for Subranges**

We introduce the code for one subrange containing the solution. When we get enough quantum computing time, we will update the general code. This code produces code that can run a linear system on a D-Wave quantum annealer.

```
import numpy as np
import random, math
```

```python
import copy
import sys
import math

Dimension = 32
A = np.random.randint(-10, 10, size=(Dimension,Dimension))
np.set_printoptions(threshold=sys.maxsize)
print("Matrix A:")
print(A)

x_fin = np.random.randint(-128, 127, size=Dimension)
print("\nSolution vector x:")
print(x_fin)

b = np.dot(A, x_fin)
print("\nVector b:")
print(b)

qubits = 2
# Calculate translation number including solution vector x
Trs = np.zeros(Dimension)
x_qubit = np.zeros(Dimension)
sub_size = pow(2,qubits)
for k in range(Dimension):
    Val = divmod(x_fin[k], sub_size)
    Trs[k] = sub_size*Val[0]
    x_qubit[k] = Val[1]
print("Translation vector T:")
print(Trs)
print("\nQubit notation:")
print(x_qubit)
c = b - np.dot(A, Trs)
print("\nUpdated vector c:")
```

```python
print(c)

QM = np.zeros((qubits*Dimension, qubits*Dimension))
### Linear terms ###
for k in range(Dimension):
    for i in range(Dimension):
        for l in range(qubits):
            cef1 = pow(2,2*l)*pow(A[k][i],2)
            cef2 = pow(2,l+1)*A[k][i]*c[k]
            po = qubits*i + l
            QM[po][po] = QM[po][po] + cef1 - cef2

### First quadratic term ###
for k in range(Dimension):
    for i in range(Dimension):
        for l1 in range(qubits-1):
            for l2 in range(l1+1,qubits):
                qcef = pow(2, l1+l2+1)*pow(A[k][i],2)
                po1 = qubits*i + l1
                po2 = qubits*i + l2
                QM[po1][po2] = QM[po1][po2] + qcef

### Second quadratic term ###
for k in range(Dimension):
    for i in range(Dimension-1):
        for j in range(i+1,Dimension):
            for l1 in range(qubits):
                for l2 in range(qubits):
                    qcef = pow(2, l1+l2+1)*A[k][i]*A[k][j]
                    po1 = qubits*i + l1
                    po2 = qubits*j + l2
                    QM[po1][po2] = QM[po1][po2] + qcef
```

```python
# Print Matrix Q
print("# Matrix Q is")
print(QM)
print("\nMinimum energy is ",-np.dot(c,c))
print("\n")

# Print Python code for the run in D-Wave quantum processing unit
print("Running code for D-Wave:\n")
print("from dwave.system import DWaveSampler, EmbeddingComposite")
print("sampler_auto = EmbeddingComposite(DWaveSampler(solver={'qpu': True}))\n")
print("linear = {", end = "")
for i in range(qubits*Dimension-1):
    linear = i + 1
    print ("('q",linear,"','q",linear,"'):",format(QM[i][i]),sep='', end = ", ")
print ("('q",qubits*Dimension,"','q",qubits*Dimension,"'):",format(QM[qubits*Dimension-1][qubits*Dimension-1]),"}", sep='')

print("\nquadratic = {", end = "")
for i in range(qubits*Dimension-1):
    for j in range(i+1,qubits*Dimension):
        if QM[i][j] != 0:
            qdrt1 = i + 1
            qdrt2 = j + 1
            if i == qubits*Dimension-2 and j == qubits*Dimension-1:
                print ("('q",qdrt1,"','q",qdrt2,"'):",format(QM[i][j]), "}", sep='')
            else:
                print ("('q",qdrt1,"','q",qdrt2,"'):",format(QM[i][j]), sep ='', end = ", ")

print("\nQ = dict(linear)")
print("Q.update(quadratic)\n")

qa_iter = 1000
```

```
print("sampleset = sampler_auto.sample_qubo(Q, num_reads=",qa_iter,")", sep =
"")
print("print(sampleset)")
```

## Discussion

Since the entire range is divided into subranges and only $m + 1$ qubits are used to represent each $x_i$ in each subrange, it is simple to calculate the QUBO formula. When the position of the subrange is changed, the changed part of the new QUBO formula is $-2\vec{c}^T A\vec{x}$ and only the linear term can be calculated using the new constant vector $\vec{c}$. For each subrange, QUBO formula is $\vec{x}^T A^T A \vec{x} - 2\vec{c}^T A\vec{x}$ and it becomes a solution when it has the minimum value $-\vec{c}^T\vec{c}$ (see Eq. 14).

The following example is the subrange in which the solution of $\|A\vec{x'} - \vec{b}\|_2^2$ exists when $T_{1,s} = 16$ and $T_{2,s} = -32$. Here, $A = \begin{pmatrix} 3 & 1 \\ -1 & 2 \end{pmatrix}$ and $\vec{b} = \begin{pmatrix} 46 \\ -55 \end{pmatrix}$. We represent $x_1'$ and $x_2'$ by using 8 qubits, $T_{1,s}$ and $T_{2,s}$. Then

$$x_1' = q_1 + 2q_2 + 4q_3 + 8q_4 + 16,$$

$$x_2' = q_5 + 2q_6 + 4q_7 + 8q_8 - 32,$$

and the QUBO formulation $Q'$ for the subrange is

$$Q' = \begin{pmatrix} -120 & 40 & 80 & 160 & 2 & 4 & 8 & 16 \\ 0 & -220 & 160 & 320 & 4 & 8 & 16 & 32 \\ 0 & 0 & -360 & 640 & 8 & 16 & 32 & 64 \\ 0 & 0 & 0 & -400 & 16 & 32 & 64 & 128 \\ 0 & 0 & 0 & 0 & -155 & 20 & 40 & 80 \\ 0 & 0 & 0 & 0 & 0 & -300 & 80 & 160 \\ 0 & 0 & 0 & 0 & 0 & 0 & -560 & 320 \\ 0 & 0 & 0 & 0 & 0 & 0 & 0 & -960 \end{pmatrix}.$$

On the other hand, the QUBO formula $Q$ for original $\vec{x}$ is as below.

$$Q = \begin{pmatrix} -3760 & 40 & 80 & 160 & 2 & 4 & 8 & 16 \\ 0 & -732 & 160 & 320 & 4 & 8 & 16 & 32 \\ 0 & 0 & -1384 & 640 & 8 & 16 & 32 & 64 \\ 0 & 0 & 0 & -2448 & 16 & 32 & 64 & 128 \\ 0 & 0 & 0 & 0 & -133 & 20 & 40 & 80 \\ 0 & 0 & 0 & 0 & 0 & 276 & 80 & 160 \\ 0 & 0 & 0 & 0 & 0 & 0 & 592 & 320 \\ 0 & 0 & 0 & 0 & 0 & 0 & 0 & 1344 \end{pmatrix}.$$

This example shows that only the linear term changes when calculating QUBO formula by changing the subranges. When using 2 qubits for each $x_i$ for a 4 by 4 matrix, the total occurrence of all combination of qubits having the lowest energy is 545 from 1,000 anneals. For a 16 by 16 matrix, the total occurrence of all combination of qubits having the lowest energy is 1 from 50,000 anneals. For a 32 by 32 matrix, the total occurrence of all combination of qubits having the lowest energy is 0 from 100,000 anneals. Now, we need to do a lot of annealing for a large matrix to get a solution. However, this is not a big deal since we know the minimum the QUBO model for each subregion should have. Also, the eigenvalue problem can be solved in the similar way.

"Example codes for our results: https://github.com/ktfriends/Numerical_Quantum_Computing"